\documentclass[reprint,prd,onecolumn,notitlepage,11pt,nofootinbib]{revtex4-1}
\usepackage[english]{babel}
\usepackage{amsmath,amssymb,graphicx,bm,lipsum,float}
\usepackage[colorlinks=true,citecolor=blue,linkcolor=blue,urlcolor=blue]{hyperref} 
\usepackage[font={small},flushleft,indent]{caption}
\usepackage{setspace}

\begin{document}

\title{Auto- and cross-correlations for multiple images of corotating hotspots in accretion disks}
\author{Qing-Hua Zhu}
\email{zhuqh@cqu.edu.cn} 
\affiliation{School of Physics, Chongqing University, Chongqing 401331, China} 
 
\begin{abstract} 
Due to the short gravitational timescale of Sgr A*, variable emissions near the galactic center are expected in the Very-long-baseline interferometry observations. Phenomenologically, the high-flux variable emissions could be interpreted as occasional events from hotspots within accretion disks.  It provides a probe of black hole (BH) geometry and accretion matter in the strong-field
regime of gravity. In this study, we find that light curve profile alone is not proper for distinguishing BH geometries, as our results show that the profiles, even including those from higher-order images, are dependent on hotspot shapes, which are known in practice as amorphous. To alleviate this situation, we examine the spatial-temporal correlations between multiple images of the corotating hotspots.  Our results find that the correlations, particularly those from higher-order images, could serve as a robust observable to reflect the inclination angles and BH geometries, because i) the correlated band structure is independent of the hotspot shapes, and ii) the correlations from higher-order images could encode BH geometries and exhibit no overlap with observational signatures from the lower-order ones. We present a comprehensive study on correlations from primary the eighth-order images with various orbital configurations and inclination angles, and show its observational signatures. It is expected that BH geometries can be inferred via the spatial-temporal correlation analysis.

\end{abstract} 

\maketitle

\section{Introduction}

Recent horizon-scale observations of the supermassive black hole (BH) at our galactic center might open up a new window to study spacetime geometries and accretion matter in the strong-field regime of gravity \cite{Narayan:2023ivq,Levis:2023tpb,Galishnikova:2022mjg,EventHorizonTelescope:2022wkp,EventHorizonTelescope:2022exc,GRAVITY:2018sef,GRAVITY:2023avo}. 
Specifically, the Event Horizon Telescope (EHT) collaboration captured both time-averaged images and sequential snapshots of Sgr A*, revealing a ring-like structure at the horizon scale \cite{EventHorizonTelescope:2022wkp,EventHorizonTelescope:2022exc}. The GRAVITY collaboration has reported flare events within ten gravitational radius \cite{GRAVITY:2018sef,GRAVITY:2023avo}. These flares exhibit clockwise orbital motion on the sky and could be modeled as a single dominant hotspot in the innermost accretion zone \cite{GRAVITY:2020lpa,GRAVITY:2023avo}.

Because the gravitational timescale of Sgr A* is approximately twenty seconds, variable emissions near the black hole are expected in the observations, including variability in snapshots of the black hole images \cite{EventHorizonTelescope:2022exc} and flare motions measured via astrometry and polarization \cite{GRAVITY:2023avo}. Phenomenologically, there are two distinct scenarios for the variable emission configurations \cite{Genzel:2010zy}. The first one is stochastic fluctuations in the accretion disk \cite{Mauerhan:2005xx,2009ApJ...703.1323D,Cardenas-Avendano:2022csp}, which might physically originate from magnetohydrodynamic turbulence \cite{Chan:2006pb}. It was used to interpret red-noise variability in the near-infrared emissions \cite{Do:2008uf}. The second one is a compact flare region, modeled by orbiting hotspots in the accretion disk \cite{Broderick:2005jj,Broderick:2005my}. It provides an interpretation of quasi-periodic changes of centroid positions or linear polarizations in BH observations \cite{Genzel:2003as,GRAVITY:2018sef,GRAVITY:2023avo,Eckart:2006fc,2006A&A...460...15M,Trippe:2006jy,Zamaninasab:2009df}. There is also a hybrid model in which the variable emissions are composed of two states, known as the quiescent and flare states \cite{Dodds-Eden:2009wba,2011ApJ...728...37D}. It suggests a statistical description of low-flux emissions and occasional flare events at higher flux levels.
\cite{Genzel:2010zy}. This model predicts a tailed log-normal flux distribution, which is supported by recent observations of Sgr A* \cite{GRAVITY:2020xcu}. 

Thus, the variable emissions resulting from stochastic fluctuations and deterministic sources are both crucial for enhancing our understanding of Sgr A*. To explore properties of the emissions, the correlation analysis were used to search for periodicity of emissions near BH \cite{Berkley:2000mp,Gandhi:2008qr,Fukumura:2010bw}. Because multiple images of emission sources are intrinsically correlated, pioneers found that correlations can reflect the temporal lapse between successive images even for stochastic emissions \cite{Chesler:2020gtw,Hadar:2020fda,Qian:2021aju,Hadar:2023kau,Harikesh:2025nmt}. In this framework, the temporal variability in Sgr A* images can be attributed to stochastic sources with correlation timescale three times longer than time delay between light echoes \cite{Cardenas-Avendano:2024sgy}. Additionally, inspired by recent high resolution observations, it was extended to spatiotemporal correlations \cite{Hadar:2020fda,EventHorizonTelescope:2022okn}, in which both the critical parameters, azimuthal lapse and temporal lapse, are important. For deterministic sources, the spatiotemporal correlations was employed addressing a single orbiting hotspot \cite{Wong:2020ziu,ZhenyuZhang:2025cqn}, and pattern speed of accretion disk \cite{Conroy:2023kec}. Theoretically, the studies on deterministic emissions might be as interesting as those for stochastic emissions. 

This paper investigates the spatiotemporal auto- and cross-correlations for multiple images of an orbiting hotspot near the BH. Since higher-order images are expected to encode the BH geometries independently of emission sources, like BH shadow \cite{Narayan:2023ivq}, we are well-motivated to study the effects from higher-order images on BH image variability, and associate them with specific emission configurations, such as inclination angle of accretion disks and the hotspot shapes. While the conceptions and properties of higher-order images were well-established  \cite{Gralla:2019xty,Gralla:2019drh,Johnson:2019ljv,Wielgus:2021peu,Bisnovatyi-Kogan:2022ujt,Tsupko:2022kwi,Wang:2022mjo,Kocherlakota:2024hyq,Igata:2025xxb}, the studies on correlations for the deterministic emissions were limited to lower-order images \cite{Wong:2020ziu,Conroy:2023kec,ZhenyuZhang:2025cqn}, likely due to high computational costs. This study employs an efficient ray-tracing scheme for point-like hotspots developed in Ref.~\cite{Zhu:2024vxw}, and extensively present the correlations for multiple images, from primary to eighth-order images. This allows us to compare the difference of observational signatures between higher- and lower-order images.  In fact, it is expected that the lapses in azimuthal angle and time could be influenced by configuration of emission sources when lower-order images are considered, because the critical parameters, Lyapunov exponent, azimuthal lapse and temporal lapse, were derived for higher-order images \cite{Gralla:2019xty,Gralla:2019drh,Hadar:2020fda,Johnson:2019ljv}. Additionally, we also present the general observational signatures in the correlation. The inclination angle can affect the correlated region in the correlations, because the apparent rotation speed of a hotspot is nonuniform for a high inclination angle. The hotspot shapes can affect the apparent solid angles, thereby altering the profiles of light curves. We validate this point by adopting the toy models of hotspot as spherical compact objects or distributed on the surface of accretion disks. It shows that the hotspot shapes have little influence on the correlations. All these results indicate that the correlations might serve as a more robust observable than the light curve profiles to reflect BH geometries.

The rest of the paper is organized as follows. In Sec.~\ref{II}, we briefly review the ray-tracing scheme we employed in the subsequent sections. In Sec.~\ref{III}, we introduce the toy models of hotspot as a spherical compact object and distributed on the surface in  accretion disks. Semi-analytical formulas are derived for calculating observed intensities and fluxes of the hotspots. In Sec.~\ref{IV}, we present the results of auto- and cross correlations for multiple images of orbiting hotspots, and illustrate the observational signatures. In Sec.~\ref{V}, conclusions and discussions are summarized.

\section{Ray tracing scheme for spherical black hole \label{II}}

In this section, we briefly review the ray tracing scheme for efficiently imaging of orbiting hotspots \cite{Zhu:2024vxw}, which is based on a one-to-one mapping as follows, 
\begin{eqnarray}
  \text{RT} : \left( \textbf{x}_\text{o}, \textbf{x}_\text{s} \right) & \mapsto & (\text{orders of images}, \Phi, \Psi)~, \label{1}
\end{eqnarray}
where $\textbf{x}_\text{o}[=(r_\text{o},\theta_\text{o},\phi_\text{o})]$ and $\textbf{x}_\text{s}[=(r_\text{s},\theta_\text{s},\phi_\text{s})]$ are the location of observers and emission sources, respectively, and $(\Phi,\Psi)$ are the celestial coordinate on the sky. This scheme differs from traditional ray tracing \cite{1992MNRAS.259..569K} by projecting the emission sources directly onto observers' sky.

Specifically, for a BH metric in the form of $\mathrm{d} s^2 = - f (r) \mathrm{d} t^2 + \frac{\mathrm{d} r^2}{f (r)} + r^2 \mathrm{d} \Omega ^2$, the 4-momentum of light rays is
\begin{eqnarray}
  k_\mu = \left(-E, \pm_r\frac{1}{f} \sqrt{E^2 - \frac{K   f  }{r^2}},\pm_{\theta} \sqrt{K - \frac{L^2}{\sin^2 \theta}},L\right)~, \label{2}
\end{eqnarray}
where the $E$, $K$, and $L$ are three integral constants. These constants determine the worldlines of light rays. On the observers' sky, the celestial coordinates  can be written in the form of
\begin{subequations}
  \begin{eqnarray}
    \Phi & = &   \varphi ~,\\
    \Psi & = & \arccos 
    \sqrt{1 - \frac{\rho^2 f (r_o)}{r_o^2}}  ~,
  \end{eqnarray} \label{3}
\end{subequations}
where $\rho \equiv  {\sqrt{K}}/{E} $, $\cos \varphi  \equiv  - {L}/{(\sqrt{K} \sin \theta_\text{o})}$. By making use of Eq.~(\ref{2}), the geodesic equations can be evaluated to be
\begin{subequations}
  \begin{eqnarray}
    0 & = & \mp_{\theta} \int_{\theta_\text{s}}^{\theta_\text{o}} \frac{\mathrm{d} \theta}{\sqrt{1 -
    \csc^2 \theta \sin^2 \theta_o \cos^2 \varphi}} \pm_r \rho \int_{r_\text{s}}^{r_\text{o}}
    \mathrm{d} r \left\{ \frac{1}{r \sqrt{r^2 - \rho^2 f}} \right\}~, \\
    \phi_s - \phi_s & = & \mp_{\theta} \sin \theta_o \cos \varphi
    \int_{\theta_\text{s}}^{\theta_\text{o}} \mathrm{d} \theta \left\{ \frac{1}{\sin^2 \theta \sqrt{1 -
    \csc^2 \theta \sin^2 \theta_o \cos^2 \varphi}} \right\} ~.
  \end{eqnarray} \label{4}
\end{subequations}
These integrals are performed along the worldline of light ray, and $\pm_r$ and $\pm_\theta$ change from $\pm$ to $\mp$ when encountering a turning point of light ray. As shown in Eqs.~(\ref{4}), the $\rho$ and $\varphi$ are separated in the integrals for $r$ and $\theta$, due to spherical symmetry of BH. Associating Eq.~(\ref{3}) with Eq.~(\ref{4}), the ray-tracing scheme is established through the inverse determination of  $(\Psi,\Phi)$ based on fixed   $(r_\text{o},\theta_\text{o},\phi_\text{o})$ and $(r_\text{s},\theta_\text{s},\phi_\text{s})$. To enhance efficiency in implementing ray tracing, analytical results of Eqs.~(\ref{4}) were  employed, and these technique details were presented in Ref.~\cite{Zhu:2024vxw}.

In addition to the mapping between the two spatial points, the time experienced by the observers is given by the proper time at $\textbf{x}_\text{o}$, namely,
\begin{eqnarray}
  \tau = t_\text{o}\sqrt{f(r_\text{o})} = \sqrt{f(r_\text{o})}\left(t_\text{s}\pm_r \int_{r_\text{s}}^{r_\text{o}} \textrm{d} r \left\{ \frac{r}{f  \sqrt{r^2 - \rho^2 f}} \right\} \right)~.\label{5}
\end{eqnarray}
The second equality shows that the observed time $t_\text{o}$ is determined by emission time $t_\text{s}$ and propagation time of light ray. For the latter one, it depends on emission location $r_\text{o}$ and the parameter $\rho$. 

\begin{figure}[H]
  \centering
  \includegraphics[width=1\linewidth]{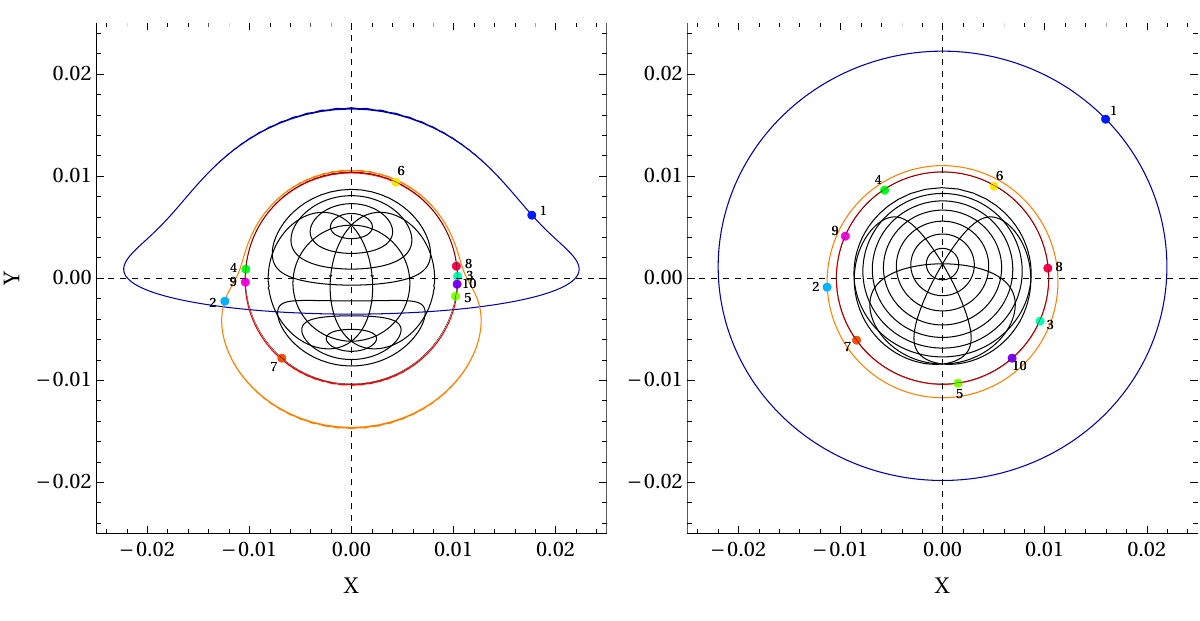} 
  \caption{The multiple images of corotating point-like hotspots for inclination angles $4\pi/9$ (left panel) and $\pi/9$ (right panel). The numbers represent the image orders. For example, the no.1 represents the primary image. The solid curves represent the apparent tracks of hotspots, and the black grids denote apparent region of the event horizon. We have introduced $(\text{X},\text{Y})\equiv(\tan\Psi \cos\Phi, \tan\Psi \sin\Phi)$, which is a gnomonic projection of the celestial coordinates $(\Phi,\Psi)$. \label{F1}}
\end{figure}

The gravitational field near BH is extremely intense, such that the bending light rays could wind round the BH multiple times. It thus leads to multiple images of emission sources. 
Figure~\ref{F1} shows the multiple images of corotating point-like hotspots for selected inclination angles. There are two noticeable facts: a) the apparent position angle $\Phi$ of these images are different, due to the lapse in azimuthal angle \cite{Gralla:2019drh}, and b) at low inclination angles, higher-order images tend to be distributed uniformly across the photon ring, whereas at high inclination angles, they mostly localize the regions of $\Phi \rightarrow 0$ or $\pm\pi$. For the latter point, it is because the apparent rotation speeds of these images are not uniform. In Sec.~\ref{IV}, we will show that these facts are also reflected in correlations for  multiple images of hotspots.


\

\section{Intensities and fluxes of point-like hotspots \label{III}}

This study focuses on individual flare event near the BH, which could be modeled as an orbiting hotspot. Specifically, we consider the hotspot as a point source in the accretion disk, and the specific fluxes are given by 
\begin{eqnarray}
  F_\nu^\text{tot} & = & \sum^{\infty}_{n = 0} \sum_{O}^{\{ \text{A}, \text{B} \}}
  F^{(  n, O)}_\nu = \sum_{n = 0}^{\infty} \sum_{O}^{\{ \text{A}, \text{B} \}} \int
  I_{\text{obs}}^{(n, O)} \left( \Psi, \Phi ; t_{\text{o}},
  \textbf{x}_{\text{o}} \right) \mathrm{d} \Omega ~,\label{6}
\end{eqnarray}
where
the $n$ represents wind-number of light path, the apparent solid angle is defined with celestial coordinates on observer's sky, namely, $\mathrm{d} \Omega  \equiv  \sin \Psi \mathrm{d} \Psi \mathrm{d} \Phi$, we have $O\in \{\text{A},\text{B}\}$ and the superscripts A and B represent A-type and B-type images classified in Ref.~\cite{Zhu:2024vxw}. Image orders from primary to eighth-order correspond to the labels $(0,\text{A})$, $(0,\text{B})$, $(1,\text{A})$...$(3,\text{B})$. The $I_{\text{obs}}$ denotes observed intensity at frequency $\nu$, and can be derived from
\begin{eqnarray}
  I_{\text{obs}} \left( \Psi, \Phi ; t_{\text{o}}, \textbf{x}_{\text{o}}
  \right) & = & g \left( \Psi,\Phi, \textbf{x}_{\text{s}},
  \textbf{x}_{\text{o}} \right)^3 I_{\text{emt}} \left( t_{\text{s}},
  \textbf{x}_{\text{s}} \right)~,
\end{eqnarray}
where the redshift factor is $g [ =  {(\left. u_{\text{o}} \cdot k   \right|_{\textbf{x}_{\text{o}}})}/{(\left. u_{\text{s}} \cdot k   \right|_{\textbf{x}_{\text{s}}})}$] and the $I_\text{emt}$ is emission intensity of a hotspot.

\begin{figure}
  \centering
  \includegraphics[width=1\linewidth]{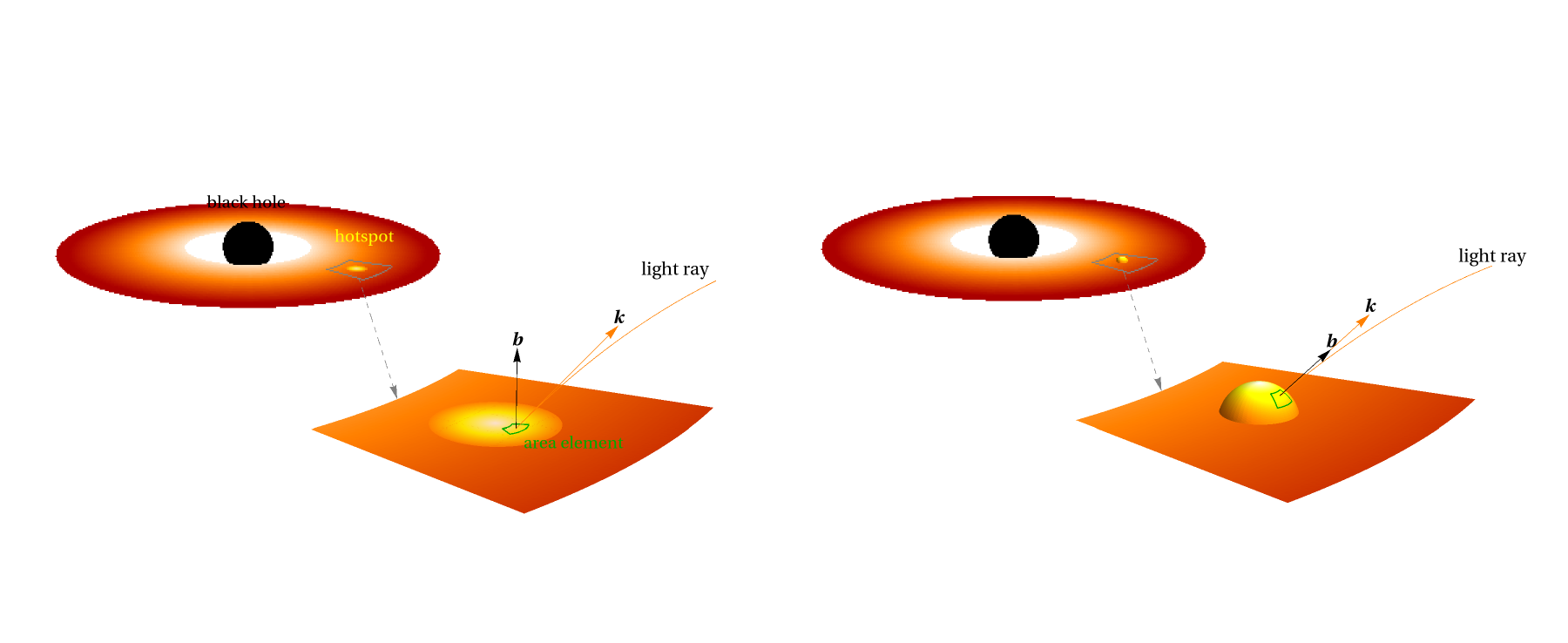}
  \caption{Schematic diagrams for emission models of hotspots. Left panel: hotspots are distributed on the surface of thin accretion disks. Right panels: hotspots are in a shape of sphere.  \label{F2}}  
\end{figure}

For each image order, denoted by $(n,O)$, all the spatial positions outside the BH can be projected onto the observer's sky \cite{Zhu:2024vxw}. In this sense, the solid angle in Eq.~(\ref{6}) can be mapped onto the surfaces of hotspots. It appears intuitive, because what we see is the outer surface of an object. By making use of Eqs.~(\ref{3}) and (\ref{4}), the Jacobi determinants can be given by
  \begin{eqnarray}
    \left| \frac{\partial \left( r_{\text{s}}, \phi_{\text{s}} \right)}{\partial
    (\Psi, \Phi)} \right| & = & \left| \frac{\partial \left( r_{\text{s}},
    \phi_{\text{s}} \right)}{\partial (\rho, \varphi)} \right| \left|
    \frac{\partial (\rho, \varphi)}{\partial (\Psi, \Phi)} \right| = \left|
    \frac{b_{\theta}}{S \cdot b} \right| \frac{r_{\text{o}}^2}{f \left(
    r_{\text{o}} \right)} \sin \Psi | \cos \Psi |~, \label{8}
  \end{eqnarray} 
where we have used the relations between $(\rho, \varphi)$ and $\left( r_{\text{s}}, \phi_{\text{s}} \right)$ based on Eqs.~(\ref{4}), namely, $  \rho \mathrm{d} \rho \mathrm{d} \varphi  =   \left| {S^{\mu} b_{\mu}}/{b_{\theta}} \right| \mathrm{d}  r_{\text{s}} \mathrm{d} \phi_{\text{s}}$, the $b_{\mu}$ is the normal vector of the two-surface, and $S^{\mu}$ is given by
\begin{subequations}
  \begin{eqnarray}
    S^t & = & 0~,\\
    S^r & = &  \left( \frac{\partial}{\partial \rho} (\rho I_r)
    \right)^{- 1} \frac{(- 1)^{l + 1} \rho \sin \theta_s (1 - \sin^2 \theta_o \cos^2
    \varphi)}{\cos \theta_o \sqrt{\sin^2 \theta_s - \sin^2 \theta_o \cos^2
    \varphi} + (- 1)^{l + 1} \cos \theta_s \sin \theta_o \sin \varphi}~,\\
    S^{\theta} & = &  \mp_r \frac{\rho^2}{r_s \sqrt{r_s^2 - \rho^2 f (r_s)}}
    \left( \frac{\partial}{\partial \rho} (\rho I_r) \right)^{- 1}
    \frac{\sqrt{\sin^2 \theta_s - \sin^2 \theta_o \cos^2 \varphi} (1 - \sin^2
    \theta_o \cos^2 \varphi)}{\cos \theta_o \sqrt{\sin^2 \theta_s - \sin^2
    \theta_o \cos^2 \varphi} + (- 1)^{l + 1} \cos \theta_s \sin \theta_o \sin
    \varphi}~,\nonumber \\ \\
    S^{\phi} & = &  \pm_r \frac{\rho^2}{r_s \sqrt{r_s^2 - \rho^2 f
    (r_s)}} \left( \frac{\partial}{\partial \rho} (\rho I_r) \right)^{- 1}
    \frac{(- 1)^l\csc \theta_s \sin \theta_o \cos \varphi (1 - \sin^2 \theta_o \cos^2
    \varphi)}{\cos \theta_o \sqrt{\sin^2 \theta_s - \sin^2 \theta_o \cos^2
    \varphi} + (- 1)^{l + 1} \cos \theta_s \sin \theta_o \sin \varphi}~,\nonumber\\ 
  \end{eqnarray}\label{9}
\end{subequations}
and $I_r\equiv \int_{r_\text{s}}^{r_\text{o}}
    \mathrm{d} r /(r \sqrt{r^2 - \rho^2 f})$.
The Jacobi determinants depend on parameter $l$ for given image order, as defined in Ref.~\cite{Zhu:2024vxw}.
The $S^\mu$ is defined with $\rho\textrm{d}\rho\wedge\textrm{d}\phi=S^r\textrm{d}\theta_\text{s}\wedge\textrm{d}\phi_\text{s}+S^\theta \textrm{d}\phi_\text{s}\wedge\textrm{d}r_\text{s}+S^\phi \textrm{d}r_\text{s}\wedge\textrm{d}\theta_\text{s}$, where $\wedge$ is wedge product. It describes a differential 2-form on the emission surface, which is ray-traced to corresponding surface element in the parameter space $(\rho,\varphi)$.   The normal vector $b_{\mu}$ depends on shapes of emission sources. Specifically, for a hotspot distributed on the surface of a thin accretion disk, we have $b_{\mu} = \left( 0, 0, r_{\text{s}}, 0 \right)$. In this case, Eq.~(\ref{8}) reduces to the results in previous studies \cite{Zhu:2024vxw,1992A257594B}. A hotspot in a shape of sphere yields $b_{\mu} = \gamma^{\ast} k_\mu / | \gamma^{\ast} k |$, where $\gamma^{\mu}_{\nu} [\equiv \delta^{\mu}_{\nu} + u^{\mu}_{\text{stc}} u_{\mu, \text{stc}}]$ is the spatial projection operator, and $u_{\nu,\text{stc}}[=-\sqrt{f}]$ is the 4-vector adapted to coordinate time $t$. The schematic diagrams for illustrating these emission models are presented in Figure~\ref{F2}.
For consistency, we consider the hotspots to be optically thick and the accretion disk to be optically thin. In this case, the dominant emissions come from the outer surface of a hotspot. And the surface emission intensity of a point-like hotspot can be given by
\begin{eqnarray}
  I_{\text{emt}} \left( \textbf{x}_{\text{s}} \left( t_{\text{s}} \right)
  \right) & \equiv & \int \mathrm{d} \sigma \int \mathrm{d} \bar{\tau} \left\{
  \frac{J}{\sqrt{- g}} \delta \left( t_{\text{s}} - t_p (\bar{\tau}) \right)
  \delta^3 \left( \textbf{x}_{\text{s}} - \textbf{x}_p (\bar{\tau}) \right)
  \right\} \nonumber \\
  & = & \left. \frac{J}{u^0 | b^{\theta} | r_{\text{s}}^2 \sin
  \theta_{\text{s}}} \delta \left( r_{\text{s}} - r_p \left( t_{\text{s}}
  \right) \right) \delta \left( \phi_{\text{s}} - \phi_p \left( t_{\text{s}}
  \right) \right) \right|_{\theta_{\text{s}} = \theta_p \left( t_{\text{s}}
  \right)}~,
\end{eqnarray}
where the $J$ is intrinsic coefficient originating from the emission sources, the $\bar{\tau}$ is proper time of the hotspot, $\sigma$ is arc-length parameter along hotspot trajectory, and the coordinate $(r_p(t),\phi_p(t))$ represent the hotspot trajectory in accretion disks. For a long-lived hotspot, one can set $J$ to be a constant. Associating the properties of dirac delta function, namely, $\delta \left( r_{\text{s}} - r_p \right) \delta \left( \phi_{\text{s}} - \phi_p \right) = \left| {\partial \left( r_{\text{s}}, \phi_{\text{s}} \right)}/{\partial (\Psi, \Phi)} \right|^{- 1} \delta (\Psi - \Psi_p) \delta (\Phi - \Phi_p)$ with Eqs.~(\ref{8}), the observed intensity can be evaluated to be
\begin{eqnarray}
  I_{\text{obs}} \left( \Psi, \Phi ; \tau, \textbf{x}_{\text{o}}   \right) & = & \frac{1}{\sin \Psi} \delta \left( \Psi - \Psi_p \left(   \tau \right) \right) \delta \left( \Phi - \Phi_p \left( \tau   \right) \right) g \left( \textbf{x}_{\text{s}}, \textbf{x}_{\text{o}}   \right)^3 \nonumber \\
  &  & \times \left( \frac{J  }{u^0 r_{\text{s}}^2 \sin   \theta_{\text{s}}} \right) \left( \frac{f \left( r_{\text{o}}
  \right)}{r_{\text{o}}^2} \right) \left| \frac{S \cdot b}{b^{\theta}
  b_{\theta} \cos \Psi} \right|_{\textbf{x}_{\text{s}} = \textbf{x}_{\text{s}}
  (\Phi, \Psi), (\rho,\varphi) = (\sin \Psi f \left( r_{\text{o}} \right)^{- 1 / 2}
  r_{\text{o}},\Phi)}~, \label{11}
\end{eqnarray}
where $(\Psi_p, \Phi_p)$ formulates apparent track of a hotspot. The observers' proper time $\tau$ can be determined by given $t_{\text{s}}$ and $\Psi$ as shown in Eq.~(\ref{5}). From Eq.~(\ref{11}), it indicates that  a point-like emission is projected on observers' sky onto a point-like one. Using Eqs.~(\ref{11}), we obtain flux of the hotspot, namely,
\begin{eqnarray}
  F_{\nu} \left( \tau, \textbf{x}_{\text{o}} \right) & = & g \left(
  \textbf{x}_{\text{s}}, \textbf{x}_{\text{o}} \right)^3 \left( \frac{J
   }{u^0 r_{\text{s}}^2 \sin \theta_{\text{s}}} \right) \left( \frac{f
  \left( r_{\text{o}} \right)}{r_{\text{o}}^2} \right) \left| \frac{S \cdot   b}{b^{\theta} b_{\theta} \cos \Psi} \right|_{\textbf{x}_{\text{s}} =
  \textbf{x}_{\text{s}} (\Phi, \Psi), \rho = \sin \Psi f \left( r_{\text{o}}
  \right)^{- 1 / 2} r_{\text{o}}, \varphi = \Phi}~. \label{12}
\end{eqnarray}
Based on Eqs.~(\ref{11}) and (\ref{12}), and recovering the superscripts for multiple images, the observed intensities take the form of
\begin{eqnarray}
  I_{\text{obs}}^{(n, O)} \left( \Psi, \Phi ; \tau, \textbf{x}_{\text{o}}
  \right) & = & \frac{F_{\nu}^{(n, O)}}{\sin \Psi} \delta \left( \Psi - \Psi_p^{(n, O)} \left(
  \tau \right) \right) \delta \left( \Phi - \Phi_p^{(n, O)} \left( \tau
  \right) \right)~. \label{13}
\end{eqnarray}
In pioneers' study \cite{Wong:2020ziu}, the flux of hotspots can be approximately  given by $F_\nu^{(n,O)}\simeq e^{-m\gamma}F_\nu^{(0,\text{A})}$, where $m$ is the image order, and $\gamma$ is Lyapunov exponent. We do not use above approximation, because one of the focuses of our study is exploring the difference between the lower-order images and higher-order images. Here, the $F_\nu^{(n,O)}$ is calculated exactly with Eq.~(\ref{12}). 
As this study focuses on effects of higher-order images, we adopt the Schwarzschild black hole  $f(r)\equiv 1-2(r_g/r)$ for simplicity. Although our ray-tracing scheme is for point-like emissions, the normal vector $b^\mu$ and apparent solid angle in Eqs.~(\ref{8}) and (\ref{9}) have already included the information of hotspot shapes. 

\begin{figure}[H]
  \centering
  \includegraphics[width=1\linewidth]{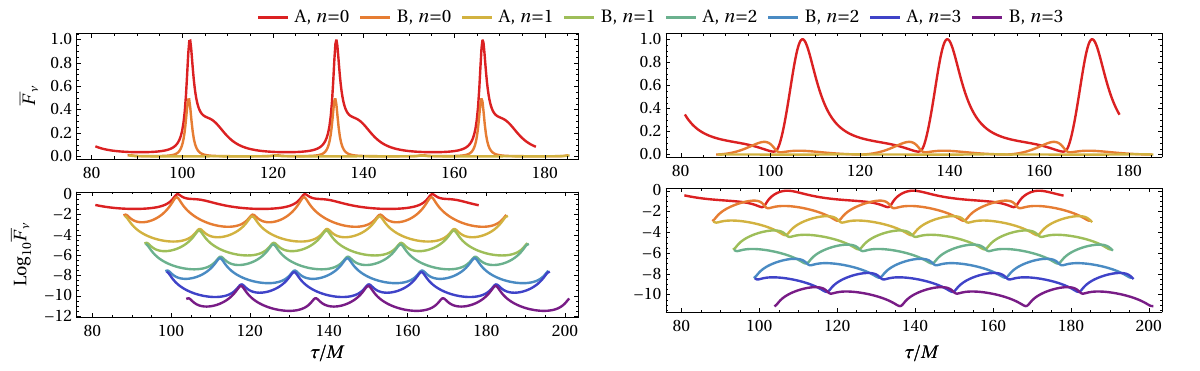}
  \caption{Normalized light curves, $\bar{F}_\nu^{(n,O)}[\equiv F_\nu^{(n,O)}/F_{\nu,\text{max}}^{(0,\text{A})}]$, from primary to eighth-order images. We consider the hotspot undergoing Keplerian motion with $r_\text{s}=10r_g$. Left panel: the emission sources distributed on the outer surface of accretion disks. Right panel: the emissions are in the shape of sphere. We consider a long-lived hotspot and set $J$ to be a constant.\label{F3}}
\end{figure}

Figure~\ref{F3} shows the normalized light curves for emission sources introduced in Figure~\ref{F2}. The results in the left panel of Figure~\ref{F3} are consistent with the those in previous studies \cite{Zhu:2024vxw,1992A257594B}. For the long-lived hotspot, the total light curves are dominated by that of primary and secondary images. Importantly, we here show that the shapes of emission sources influence the light curve profiles, because the shape can alter the apparent solid angle. And the influence also extends to the light curves from higher-order images. This feature suggests that light curve profiles are not proper to serve as an observable to infer BH geometries \cite{Li:2014coa,Li:2014fza,Rosa:2023qcv,Rosa:2022toh,Chen:2023knf}.

\

\section{Correlations for orbiting hotspots \label{IV}}

Multiple images are intrinsically correlated, because they all originate from one single emission source. In this section, we will study the correlations for multiple images of orbiting hotspots.
Utilizing Eq.~(\ref{13}), the auto-correlations of observed intensities can be given by
\begin{eqnarray}
  \mathcal{C} (\Delta t, \Delta \Phi)  = \sum_{n_1,n_2}^\infty \sum^{\text{\{\text{A},\text{B}\}}}_{X_1,X_2} \mathcal{C}^{(n_1,O_1,n_2,O_2)} (\Delta t, \Delta \Phi) ~,
\end{eqnarray}
where 
\begin{eqnarray}
  \mathcal{C}^{(n_1,O_1,n_2,O_2)} (\Delta t, \Delta \Phi) & = & \left\langle I_{\text{obs}}^{(n_1,O_1)} \left( \Psi, \Phi ; t , \textbf{x}_{\text{o}} \right) I_{\text{obs}}^{(n_2,O_2)} \left( \bar{\Psi}, \Phi + \Delta \Phi ; t + \Delta t, \textbf{x}_{\text{o}} \right) \right\rangle \nonumber \\
  & = & \int_{- \infty}^{\infty} \mathrm{d} t \int_{- \pi}^{\pi} \mathrm{d} \Phi \int \sin \Psi \mathrm{d} \Psi \int \sin \bar{\Psi} \mathrm{d} \bar{\Psi} \Big\{ \nonumber \\ && I^{(n_1,O_1)}_{\text{obs}}\left( \Psi, \Phi ; t , \textbf{x}_{\text{o}} \right) I^{(n_2,O_2)}_{\text{obs}} \left(\bar{\Psi}, \Phi + \Delta \Phi ; t + \Delta t, \textbf{x}_{\text{o}} \right) \Big\} \nonumber \\
  & = & \sum_{t_\ast} \frac{F^{(n_1,O_1)}_{\nu} \left( t_{\ast}, \textbf{x}_{\text{o}}\right) F^{(n_2,O_2)}_{\nu} \left( t_{\ast} + \Delta t, \textbf{x}_{\text{o}} \right)}{|(\Phi_p^{(n_1,O_1)} (t_{\ast}))' - (\Phi_p^{(n_2,O_2)} (t_{\ast} + \Delta t))' |}~, \label{cornn}
\end{eqnarray}
and $t_{\ast}$ is determined by equation,
\begin{eqnarray}
  \Delta \Phi + \Phi_p^{(n_1,O_1)} (t) - \Phi_p^{(n_2,O_2)} (t + \Delta t)=0~. \label{DphiDt}
\end{eqnarray} 
The $\mathcal{C}^{(n_1,O_1,n_2,O_2)} (\Delta t, \Delta \Phi)$ for $(n_1,O_1)\neq(n_2,O_2)$ represents the cross-correlation between multiple images of the hotspots, and the $\mathcal{C}^{(n,O,n,O)} (\Delta t, \Delta \Phi)$ denotes auto-correlation of multiple images.

\begin{figure}[H]
  \centering
  \includegraphics[width=0.8\linewidth]{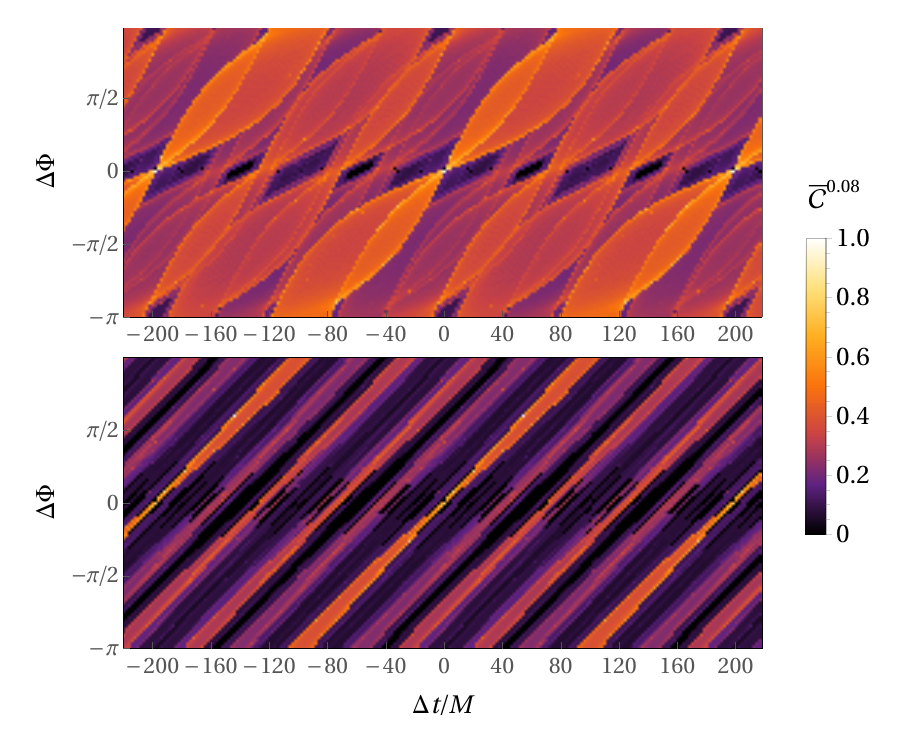} 
  \caption{\label{F4}Dimensionless auto-correlations of observed intensity, $\bar{\mathcal{C}}(\Delta t,\Delta \Phi)[\equiv{\mathcal{C}}(\Delta t,\Delta \Phi)/{\mathcal{C}}(0,0)]$. We set inclination angles to be $\pi/3$ and $\pi/9$ in the top and bottom panels, respectively. The hotspot undergoes Keplerian motion with $r_\text{s}=10r_g$. The gamma value is set to be 0.08.}
\end{figure}

\begin{figure}
  \includegraphics[width=1\linewidth]{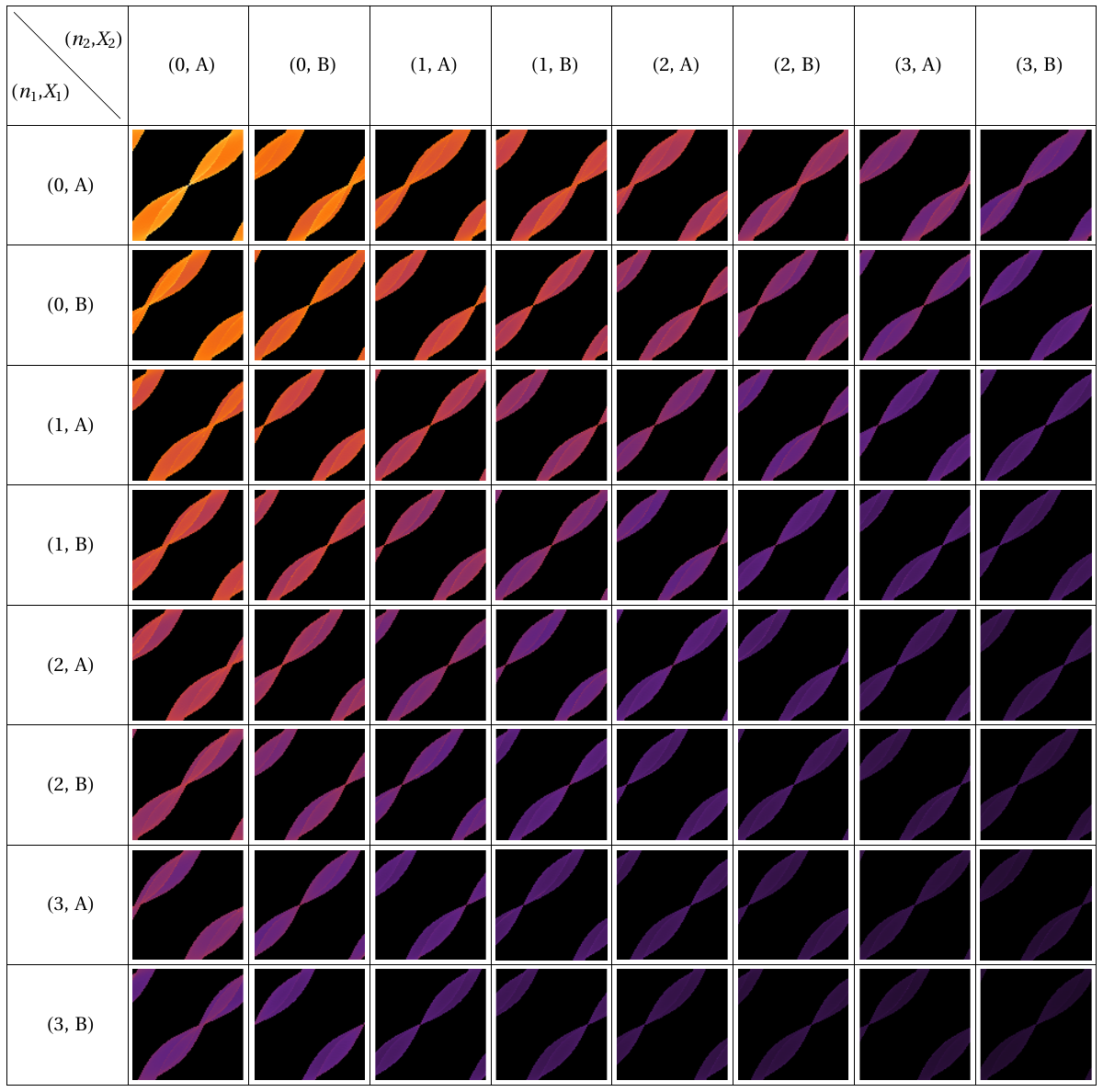}
  \caption{Dimensionless auto- and cross-correlations, $\bar{\mathcal{C}}^{(n_1,O_1,n_2,O_2)}[\equiv{\mathcal{C}^{(n_1,O_1,n_2,O_2)}}/{\mathcal{C}}(0,0)]$, for inclination angle $\pi/3$. The hotspot undergoes Keplerian motion with $r_\text{s}=10r_g$. The gamma value is set to be 0.05. \label{F5}}
\end{figure}

\begin{figure}
  \includegraphics[width=1\linewidth]{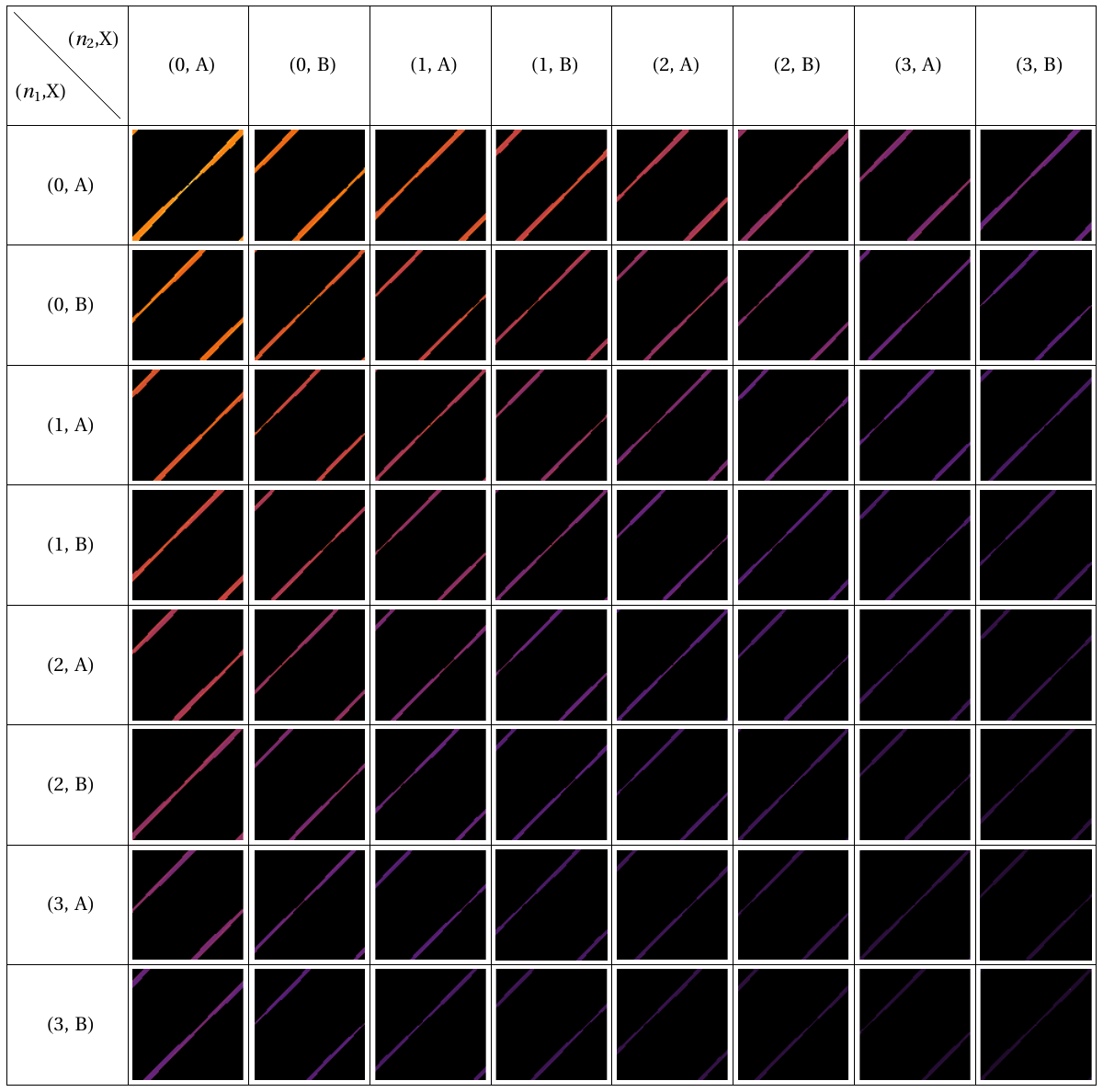}
  \caption{Dimensionless auto- and cross-correlations, $\bar{\mathcal{C}}^{(n_1,O_1,n_2,O_2)}$, for inclination angle $\pi/9$. The hotspot undergoes Keplerian motion with $r_\text{s}=10r_g$. The gamma value is set to be 0.05. \label{F6}} 
\end{figure}

Figure~\ref{F4} shows the auto-correlations of observed intensities for selected inclination angles. The auto-correlation,  $\mathcal{C} (\Delta t, \Delta \Phi)$, are the accumulation of auto- and cross-correlations for the multiple images, $\mathcal{C}^{(n_1,O_1,n_2,O_2)} (\Delta t, \Delta \Phi)$. Because of periodicity of the hotspot motions, these correlations exhibit a repeated band-like structure around $\Delta \Phi \simeq 0.03\Delta t/ M+C_0$. Here, it indicates that apparent rotation speed of the hotspot is $0.03/M$.  The bottom panel of Figure~\ref{F4} shows a narrower bandwidth in the correlations. It is because the apparent rotation speed of hotspots is nearly uniform at low inclination angles. For a high inclination angle, the apparent rotation speed decreases as the hotspot approaches the apparent angular positions $\Phi \rightarrow 0$ or $\pm\pi$. It results in a periodic modulation of the bandwidth in the correlations shown in the top panel of Figure~\ref{F4}. Comparing with the top and bottom panels, one might find that the maximum width of correlated band increases with the inclination angles. We show the auto- and cross-correlations $\mathcal{C}^{(n_1,O_1,n_2,O_2)} (\Delta t, \Delta \Phi)$ in Figures~\ref{F5} and \ref{F6}, which correspond to the cases considered in the top and bottom panels of Figure~\ref{F4}. These correlations vanish abruptly outside the bands. The instantaneous decorrelation is caused by the point-like setup of hotspots. These auto- and cross-correlations exhibit parallel band-like structures with different intercepts. The point of $(\Delta t,\Delta \Phi)=(0,0)$ lie within the correlated band of $\mathcal{C}^{(n,O,n,O)}$, because Eq.~(\ref{DphiDt}) reduces to $\Phi_p^{(n,O)}(t)-\Phi_p^{(n,O)}(t)\equiv0$. Similarly, correlated band of $\mathcal{C}^{(n_1,O_1,n_2,O_2)}$ should contain the point of $(\Delta t,\Delta \Phi)=(\delta t,\delta \Phi)$, where the $\delta t$ and $\delta \Phi$ are given by  the temporal and azimuthal lapses between multiple images \cite{Gralla:2019drh,Hadar:2020fda}. These points of $(0,0)$ and $(\delta t,\delta \Phi)$ determine intercepts of the correlated band in the correlations. 

\begin{figure}
  \includegraphics[width=1\linewidth]{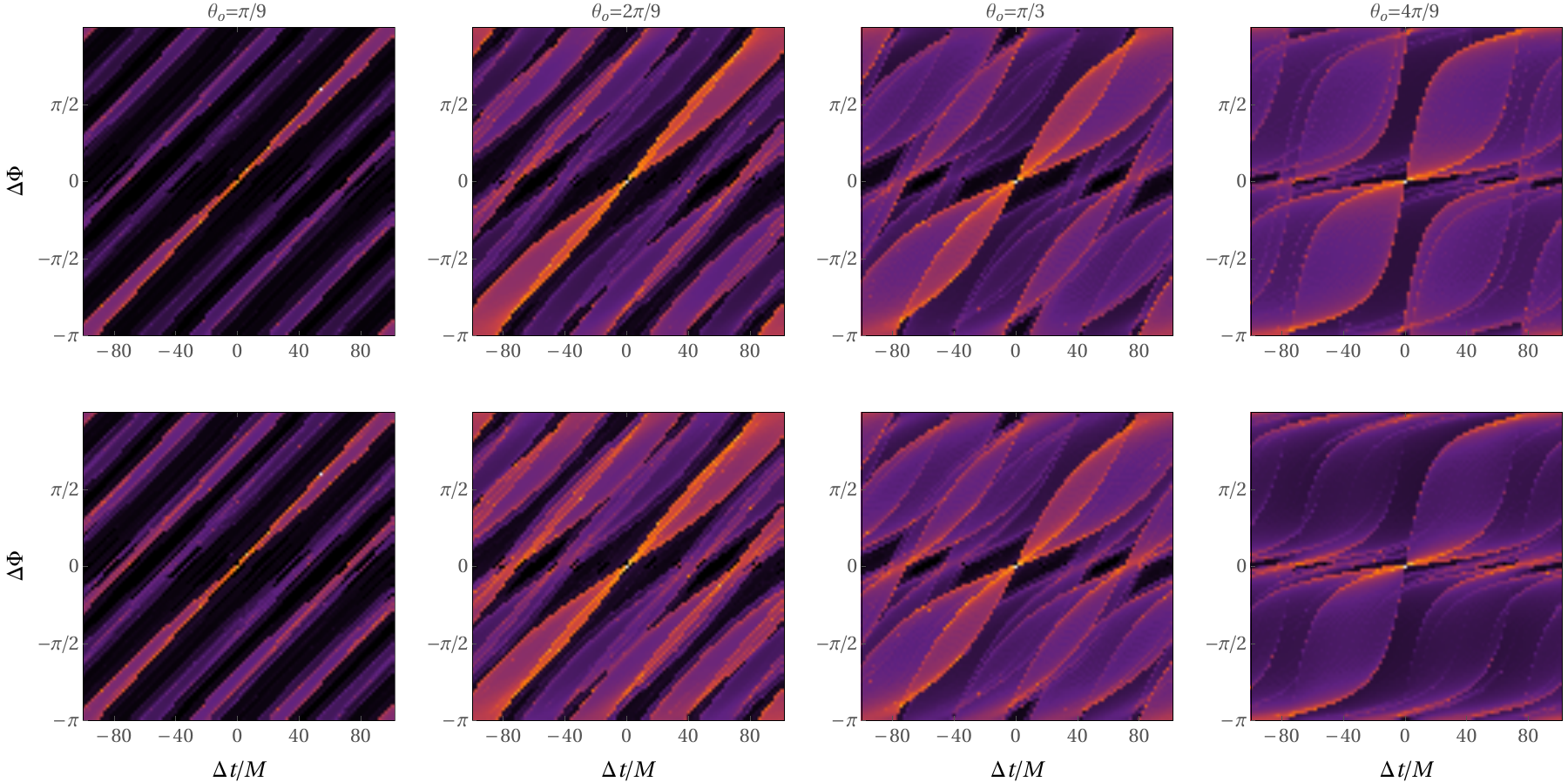}
  \caption{Dimensionless auto-correlations, $\bar{\mathcal{C}}(\Delta t,\Delta \Phi)$, for selected inclination angles from $\pi/9$ to $4\pi/9$. We consider the emission sources distributed on the outer surface of accretion disks and the emissions sources in the shape of sphere in the top and bottom panels, respectively. The gamma value is set to be 0.13.\label{F7}}
\end{figure}

From Figure~\ref{F3}, the light curves are influenced by the hotspot shapes, significantly. Thus, it is motivated to investigate whether it also can influence the correlations. We present the correlations for emission models considered in Sec.~\ref{III} in Figure~\ref{F7}. The amplitude of the correlations for the surface emissions is slightly larger than those for spherical emissions as inclination angle approaches $\pi/2$. Interestingly, the shapes of correlated bands are not influenced by the emission models. This conclusion also can be found in Eq.~(\ref{cornn}), where the correlations are dominated by the factor $|(\Phi_p^{(n_1,O_1)} (t_{\ast}))' - (\Phi_p^{(n_2,O_2)} (t_{\ast} + \Delta t))'|^{-1}$ independent of emission models.  Additionally, we once again show that the maximum width of correlated band increase with the inclination angles.

\begin{figure} 
  \includegraphics[width=1\linewidth]{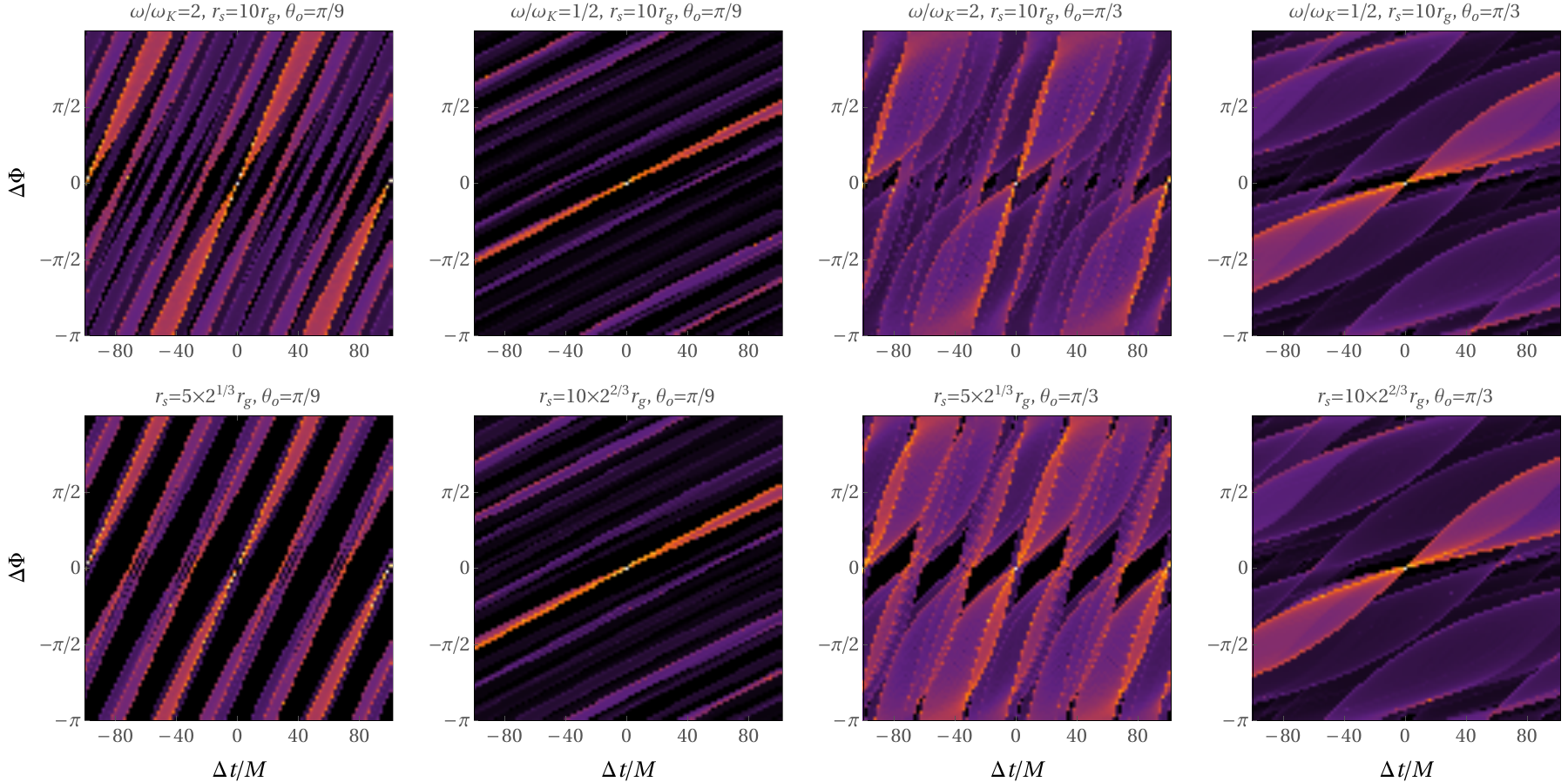}
  \caption{Dimensionless auto-correlations, $\bar{\mathcal{C}}(\Delta t, \Delta \Phi)$, for given $\omega/\omega_K$, orbital radii, and inclination angles. We consider non-Keplerian motion and Keplerian motion of hotspots in the top and bottom panels, respectively.  The gamma value is set to be 0.13. \label{F8}}
\end{figure}

\begin{figure}
  \includegraphics[width=1\linewidth]{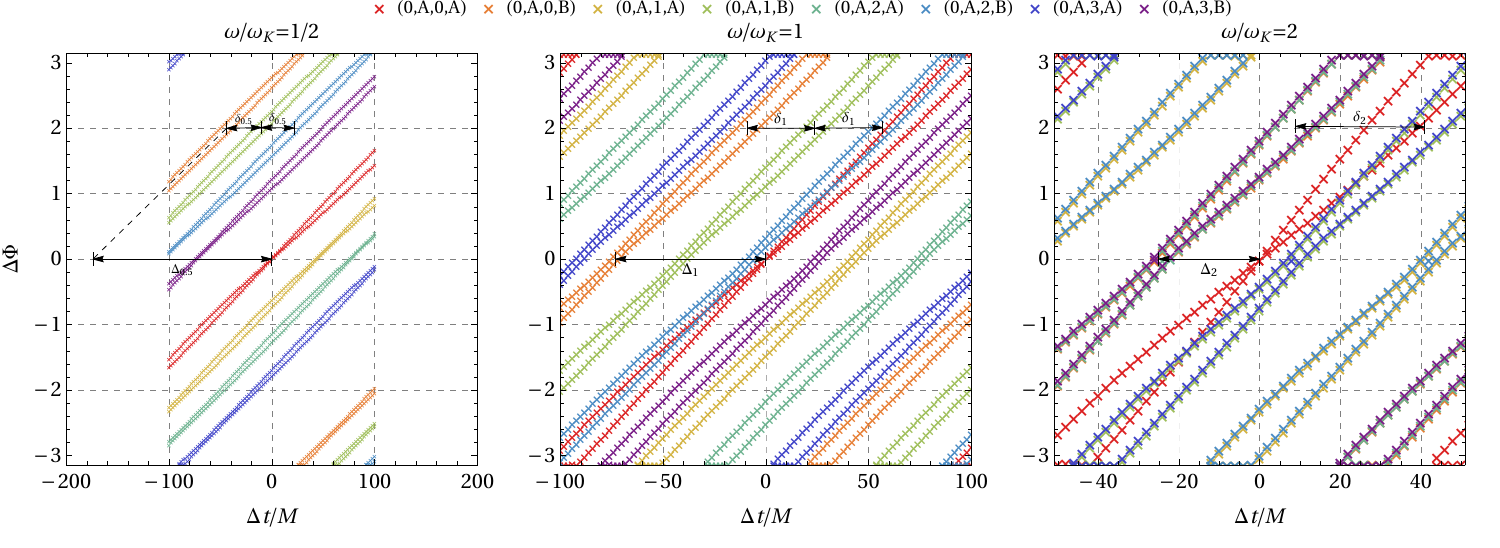}
  \caption{Correlated bands from primary-primary to primary-eighth correlations for given $\omega/\omega_K$. Here, we set $r_\text{s}=10r_g$ and $\theta_\text{o}=\pi/9$. It shows the band spacings $\delta_{0.5}\approx\delta_1\approx\delta_2$, and that the $\Delta_{\omega/\omega_K}$ decreases with $\omega/\omega_K$. \label{F9}}
\end{figure}

We also examine whether orbital configurations of hotspots influence the correlations. In Figure~\ref{F8}, we present auto-correlations for the hotspots with various rotation speeds and orbital radii. The hotspots do not necessarily undergo Keplerian motions \cite{GRAVITY:2018sef,Matsumoto:2020wul}. The slopes of the correlated bands are determined by the apparent rotation speeds of the hotspots, independent of whether the motion is Keplerian. The degree of non-Keplerian motions can be formulated by $\omega/\omega_k$, where $\omega$ and $\omega_k$ are the orbital rotation speed and Keplerian rotation speed, respectively. The top and bottom panels of Figure~\ref{F8} show the correlations for hotspots in the same rotation speeds but different orbital radii.  It is found that the spacing between correlated bands tends to be narrower for  super-Keplerian motions. 
To investigate the band spacings, we show the correlated bands for given $\omega/\omega_K$ in Figure~\ref{F9}. 
The band spacing between primary-primary and primary-secondary correlations, denoted as $\Delta_{\omega/\omega_K}$, decreases with $\omega/\omega_K$, while the band spacing between $n$th and $(n+2)$th order correlations, such as those between (0,A,0,B) and (0,A,1,B) or between (0,A,1,B) and (0,A,2,B) correlations, denoted as $\delta_{\omega/\omega_K}$, are independent of the $\omega/\omega_K$, namely, $\delta_{0.5}\approx\delta_1\approx\delta_2$. This feature indicates that the lower-order correlated bands can overlap with the higher-order ones with specific orbital configuration. We numerically find the overlap condition as follows,
\begin{eqnarray}
  \omega_\text{c}/\omega= \text{integer number}~, \label{17}
\end{eqnarray} 
where $\omega_\text{c}$ is the angular frequency of light ray on the photon sphere. As shown in the right panel of Figure~\ref{F9}, we have $\omega_\text{c}/\omega\approx3$, and thus the (0,A,0,B), (0,A,2,A) and (0,A,3,B) correlations overlap each other. The bottom-left panel of Figure~\ref{F8} exhibits the same overlap feature as the top-left one, corresponding to the right panel of Figure~\ref{F9}. The difference arises because the $\Delta_2$ is also influenced by radial distance of the hotspot, $r_\text{s}$. Due to naturality, the orbital frequency of hotspot might not satisfy the condition in Eq.~(\ref{17}). It is expected that correlations between multiple images all could be found in the $\Delta\Phi$-$\Delta t$ plots.

\begin{figure}
  \includegraphics[width=1\linewidth]{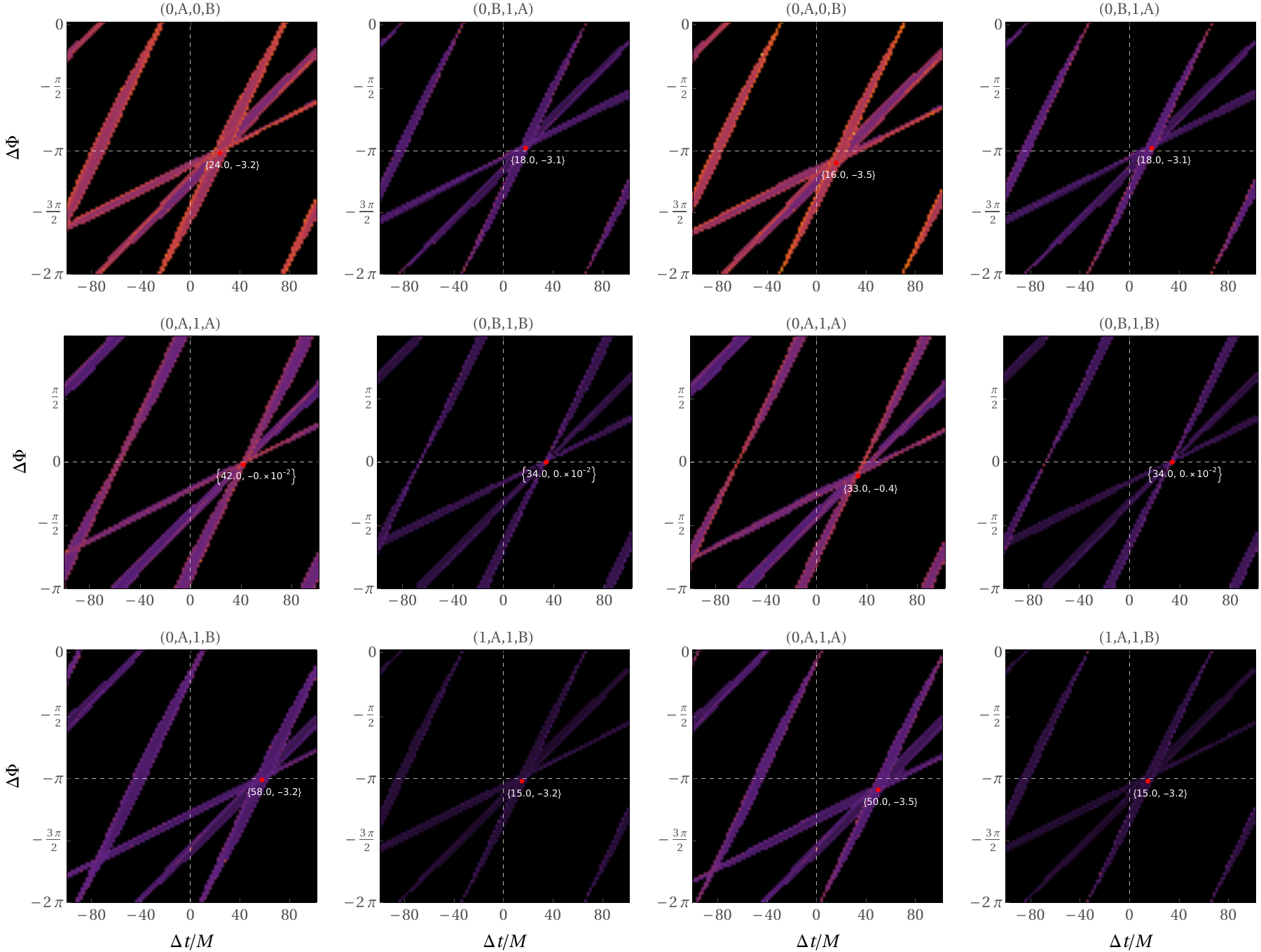}
  \caption{Dimensionless cross-correlations for multiple images for various rotation speeds of hotspots. Panels in columns 1-2: rotation speed adjusted via radii of Keplerian orbits. Panels in columns 3-4: rotation speed adjusted via non-Keplerian parameters with fixed orbital radii. The gamma value is set to be 0.1. \label{F10}}
\end{figure}

\begin{figure}
  \includegraphics[width=1\linewidth]{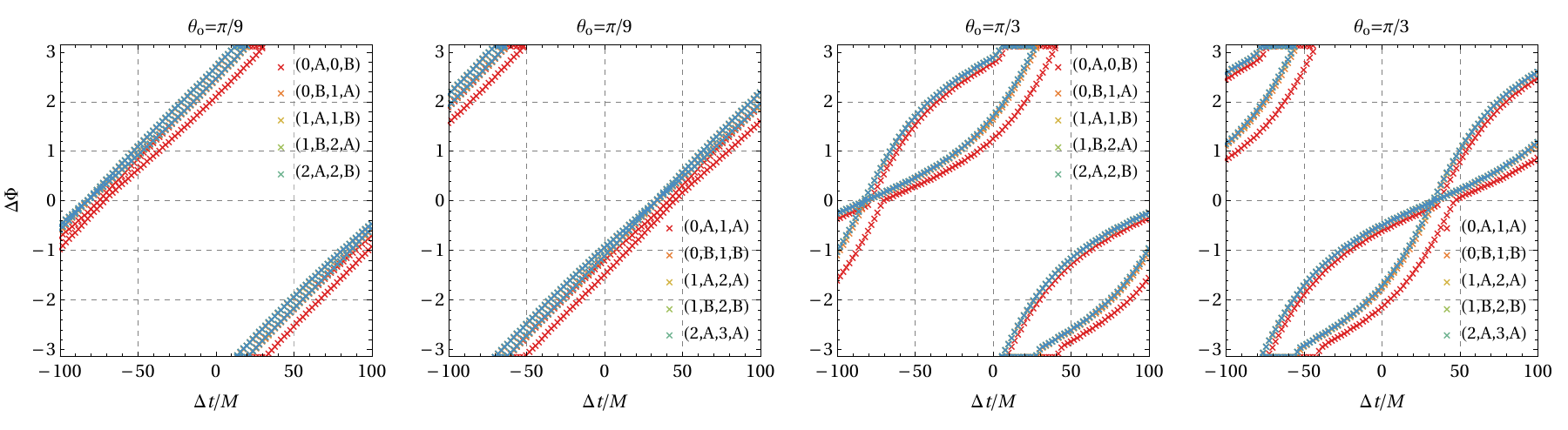}
  \caption{Correlated bands of cross-correlations between $n$th and $(n+1)$th order images (panels in columns 1 and 3) and between $n$th and $(n+2)$th order images (panels in columns 2 and 4), for selected inclination angles. \label{F11}}
\end{figure}

Because the slopes and the intercepts of correlated bands are determined by the apparent rotation speeds and lapses in azimuthal angle and time. The former is based on orbital configuration of hotspots, whereas the latter is expected to be determined by BH geometries \cite{Conroy:2023kec,Hadar:2020fda}. Therefore, one can obtain a fixed point of the correlated bands by simulating hotspots at various rotation speeds. In Figure~\ref{F10}, we present the cross-correlations for hotspots at different rotation speeds from lower- to higher-order correlations, and examine two distinct ways to change the apparent rotation speed of hotspots. The first is to adjust radii of Keplerian orbits shown in columns 1-2 of Figure~\ref{F10}, and the second is to alter non-Keplerian parameters with fixed orbital radii shown in the columns 3-4. In all cases, the correlated bands with different slopes intersect at a single point, referred as to the fixed point. As the image order increases, the angular positions of the fixed points tend to be $0$ or $\pi$.  Comparing the subplots in columns 1-2 with those in columns 3-4, the positions of fixed points are shifted for the cross-correlations involving primary images, denoted as $(0,\text{A},\ast,\ast)$ correlations, whereas the position shift is not obvious for the cross-correlations for secondary or higher-order images, denoted as $(0,\text{B},\ast,\ast)$ or $(1,\text{A},\ast,\ast)$ correlations. The shift in the fixed points is another evidence suggesting the different band spacing $\Delta_2$ in left panels of Figure~\ref{F8}. This indicates that the lapse in azimuthal angle and time between lower-order images (e.g. between the primary and secondary images) are influenced by orbital configurations of emission sources. We further show this point in Figure~\ref{F11}. The intercepts of $(0,\text{A},\ast,\ast)$ correlations are different from the higher-order ones. For higher-order images, the fixed points are solely determined by black hole (BH) parameters, because the hotspot configurations do not affect the results in the correlations. 

\ 

\section{Conclusions and Discussions\label{V}}

This paper investigated the spatiotemporal auto- and cross-correlations for multiple images of a long-lived, point-like orbiting hotspot near the BH. We extensively presented the correlations for multiple images from the primary to eighth-order. The correlations exhibit a repeated inclined band-like structure in the $\Delta \Phi$-$\Delta t$ plot.  We showed that there is a periodic modulation of width of the correlated bands, and the maximum band width increases with the inclination angles of accretion disks due to nonuniform motion of the hotspot images.  For correlations between lower-order images, the slope-intercept  of the correlated bands is determined by the azimuthal and temporal lapses of multiple images, as well as the apparent rotation speed of hotspots, consistent with previous studies \cite{Wong:2020ziu,Conroy:2023kec,ZhenyuZhang:2025cqn}. By comparing correlations for lower-order with those for higher-order images, we showed that azimuthal and temporal lapses for lower-order images is additionally dependent on orbital configuration of the hotspots. It indicates that inferring BH geometries via the correlations between lower-order images (e.g., primary-secondary cross-correlations) is less robust than those between higher-order images.

We showed that the hotspot shapes can change the apparent solid angle on the sky, consequently influencing the profiles of light curves, even including those from higher-order images.  In this sense, it is not proper to utilize the light curve profiles to infer BH geometries \cite{Li:2014coa,Li:2014fza,Rosa:2023qcv,Rosa:2022toh,Chen:2023knf}, because emission sources might be amorphous \cite{Levis:2023tpb}. 
Fortunately, although the emission shape can significantly influence light curves, our results showed that it does not change the correlated band structure of the correlations. And the higher-order cross-correlations are not overlapped with the  lower-order ones (shown in Figure~\ref{F9}). Thus, the correlations for multiple images could serve as a robust observable to reflect BH geometries in the studies of variability in BH images.


\ 

{\it Acknowledgments.} The author thanks Dr.~Zhenyu Zhang for useful discussions at the early stage of this work. This work is supported by the National Natural Science Foundation of China under grants No.~12305073 and No.~12347101.

\bibliography{cite}

\end{document}